# *Servo-based light shutters with Arduino control*

**Authors**
*Mathias S. Fischer*
*Martin C. Fischer **

**Affiliations**
*Mathias Fischer: North Carolina State University, Raleigh, North Carolina, USA*
*Martin Fischer: Department of Chemistry, Duke University, North Carolina, USA*

**Corresponding author's email address**
*Martin.Fischer@duke.edu*

**Abstract**
*In optical experiments, shutters are devices that open or close a path of light. They are often used to limit the duration of light exposure onto a target or onto a detector in order to reduce possible light-induced damage. Many commercial shutters are available for different applications – some provide very fast opening and closing times, some can handle large optical powers, and others allow for fail-safe operation. Many of these devices are costly and offer limited control options. Here we provide an open-source design for a low-cost, general purpose shutter system based on ubiquitous servo motors that are connected to an Arduino-based controller. Several shutters can be controlled by one controller, further reducing system cost. The state of the shutters can be controlled via a display built into the controller, by serial commands via USB, or by electrical control lines. The use of a microcontroller makes the shutter controller adaptable – only control options that are used need to be included, and the design accommodates a selection of display and servo options. We provide designs for all required components, including 3D print files for the servo holders and cases, software for the Arduino, libraries for serial communication (C and python), and example graphical user interfaces for testing.*

**Keywords**
*Optical shutter, Laser safety, Microscopy, Arduino, Open hardware*

| Open source license | *CERN-OHL-W-2.0 (CERN weakly reciprocal v2)* |
|---|---|
| Cost of hardware | *4 shutters with controller: ~$120* |
| Source file repository | *GitHub repo link: https://github.com/MCFLab/Shutter* |

# 1. Hardware in context

In many studies that involve a light source, light exposure is intermittent with "on"- and "off"-periods. Switching could be required for safety reasons (turning the light off while observing the sample), to limit the amount of deposited energy, to switch between different light paths, or to observe a time-dependent response. Repeatedly turning the light source on or off is often too slow (as with halogen lamps) or detrimental to the source (as with some mercury lamps). Mechanical shutters provide a convenient route to control the exposure window by inserting a mechanical block into the light path.

Many designs are in use for the type of mechanical blade, the way the blade is moved into and out of the light path, and the means of controlling the open/closed state. Common blade designs are irises (diaphragms) that open/close radially, or rigid blades that sweep across the path linearly. The blades can block the light by absorbing it (e.g., absorption on dark blades) or by redirecting it (e.g., scattering or reflection on bare metallic blades). The most common options for moving the blades are to attach them to solenoids or to rotary motors. Because of the mechanical motion, there is a tradeoff between the size of the obstructed light path and the time required to open/close it (and the frequency of opening/closing cycles). In some applications, the opening and closing times can be shortened by reducing the size of the light path by focusing and placing the shutter at the focal position. The use of a continuously rotating shutter wheel can offer vastly increased cycle frequencies, but this approach is restricted to cases where opening and closing occurs at regular intervals. Our design is not aimed at high-speed applications, but at the most common scenarios where the shutter merely provides a means to turn light on and off at arbitrary times.

Here we provide a brief overview of the available commercial shutters, which cover different applications areas. Devices with very fast opening/closing times and low latencies are available from Vincent Associates [1] or Thorlabs [2], but a single shutter with the associated driver electronics generally costs in excess of a thousand US dollars. Lower-cost devices are available that sacrifice some performance, mostly opening/closing speeds, or control options. Solenoid-based shutters are available from Brandstrom Instruments [3], EOPC [4], DACO Instruments [5], and KENDRION [6] with a wide range of blade options; Picard Instruments [7] offers a stepper-motor based shutter. Some of these devices are available with controllers and/or a programming interface. Radiant Dyes [8] offers a device based on a servo motor and a controller that opens/closes the shutter via manual switches, digital inputs, or a serial communications port. However, all these lower-cost devices still cost several hundred dollars to operate a single shutter and configuration and control options are very limited.

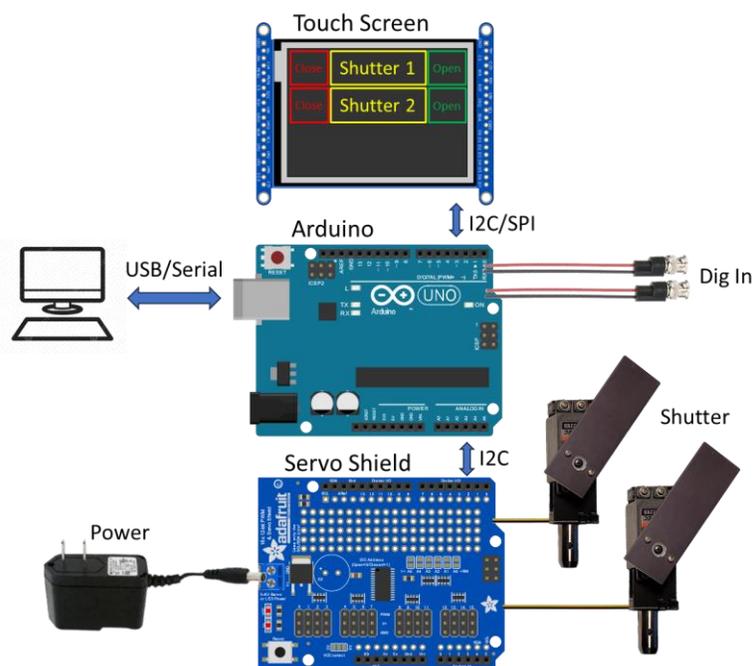

**Fig. 1.** Schematic overview of the shutter system.

Many do-it-yourself (DIY) shutter designs have been published in scientific literature. They vary greatly in performance, cost, and ease of assembly; few provide easy-to-replicate build instructions. Early designs relied on magnetic coils, relays, or solenoids [9-11], DC motors [12, 13], or loudspeakers [14]. Faster blade motion with lower timing jitter can be achieved with voice coils derived from hard disk drives [15, 16] or with a piezo cantilever design [17-19]. A simple servo-based shutter has been demonstrated [20] as part of a (only partially completed) spectrometer design, and in a quantum optics lab [21], but no details on the implementation could be found. Of these designs, the 3D printed, DC motor-based design [12] contains part files and electronic design files that aid in replication; for the voice coil design instructional material and videos [22, 23] are available.

Here we describe a simple, easy-to-build, low-cost, and open-source shutter design based on an RC servo with an Arduino-based controller that offers a wide range of configuration options. Several shutters can be controlled with one controller – we tested the design with four shutters, but extensions to 12 or more shutters are straightforward. Blades sizes are adaptable up the several centimeters. Small-beam opening/closing times of sub-10 ms and on/off cycle rates in the 10 Hz range are easily achievable. The shutter controller can be configured to receive input from a display (LCD with push buttons or a touch screen display), from hardware digital control inputs, through USB serial communication, or the shutters can be moved manually. We provide the control software (Arduino code, a C or python-based library for serial communication, and a python GUI) and the mechanical designs for two servo sizes and controller enclosures. We use our system to control the light paths in our laser-scanning microscope, for example to protect the sample from exposure when the acquisition is stopped, switch the light paths for different experimental conditions, and as a laser safety measure when the microscope is not acquiring data, but we are hopeful that our design will prove useful for many other applications.

## 2. Hardware description

### 2.1. Overall Implementation and Design

The shutter system consists of the mechanical part and the association control electronics. The mechanical shutter is based on commercial RC servos mounted in a 3D-printed holder. A black anodized aluminum blade is mounted on the servo horn (the rotatable portion) to block the light path for a given servo position. The associated control electronics is based on an Arduino Uno with a pulse width modulation (PWM) shield for controlling several servos and a display shield for displaying and switching the state of the shutters. Control of the position of the shutters can occur through user control on the display, serial communication, or TTL-compatible digital control inputs. The provided enclosures and the default hard/software configuration is designed for up to 4 channels, but the design can accommodate tens of shutters with trivial modifications.

### 2.2. Shutter mechanics

The blocking action of the shutter is achieved by rotating an opaque blade into the light path. Here, the blade is made from black anodized aluminum, which absorbs most of the light, exhibits some scattering, but transmits no light. The metallic blade dissipates moderate amounts of absorbed light as heat to the surrounding air. For high-intensity beams, care must be taken not to overheat the servo (horn and gear) and gradual bleaching of the dye used in the anodization process can be expected. For such cases, a mirror may be mounted on the blade to redirect, rather than absorb the light.

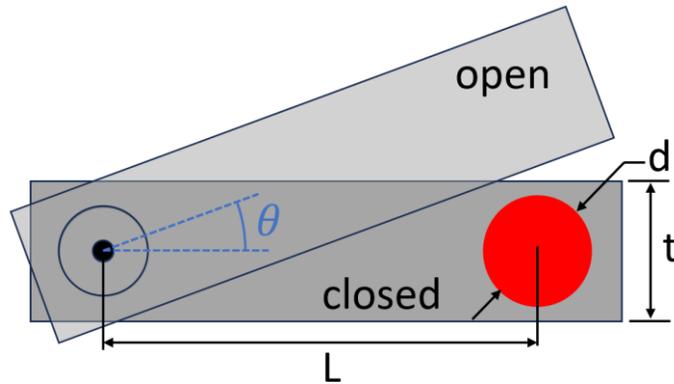

**Fig. 2.** Schematic of the shutter blade. The red circle indicates the light beam.

The speed at which the shutter opens/closes the light path depends on the rotational speed of the servo, $r$ (measured in degrees/s), and the required angle of the blade to traverse the light, $\theta$. The angle $\theta$, in turn, depends on the diameter of the light path, $d$, and its distance, $L$, from the servo axis. If $d \ll L$, the opening and closing time $\tau$ can be approximated as

$$\tau = \frac{360°}{r} \frac{d}{2\pi L} \qquad \text{Eqn. 1}$$

where we assumed a uniform beam and neglected acceleration effects of the servo. Hence, a faster servo, a longer distance between light path and servo axis, and a small light path diameter (maybe even by focusing through a lens) decreases the opening/closing times. For a laser beam of a few mm in diameter, a convenient $L$ of a few cm, and typical servo speeds (60° / 100 ms), opening and closing times are on the order of several ms. The time delay between an open/close request and the actual light increase/decrease is determined by the width of the blade, $t$, the limited acceleration of the servo, and the delay of the control electronics. Even for a minimal blade width ($t = d$) the frequency of the servo control signal limits the response to a delay of several tens of ms and timing jitter of several ms (example performance data are given below). In the accompanying material we provide part files for the mechanical holders for two servo sizes (we tested the larger size with the ubiquitous MG 996R and HiTEC HS-322; the smaller one with the faster Savöx SH-0262MG).

### 2.3. Controller

The controller for our shutter system is based on an Arduino Uno. The Arduino board can be powered through USB or via an external power supply. To drive the servos, we utilize a dedicated, external pulse width modulation (PWM) shield with a separate power supply for two reasons. First, the current that the Arduino 5V pin can supply is quite limited (800 mA if using the power input, even less if connected to an unpowered USB hub), whereas a single, medium-sized servo can already temporarily draw several hundred mA. Second, even though the servos can be PWM-controlled directly by the Arduino's digital outputs, controlling several servos requires careful sharing of the Arduino's resources (especially timers). Hence, using a dedicated, external PWM board provides enough current for many servos and significantly simplifies the programming. An optional display board serves to display the status of the shutters and to provide a way for user input. The user can also control the shutters via serial commands over USB or via external TTL control inputs monitored by the Arduino. The configuration of the shutters (such as open/close positions, labels, and control port mapping) is stored in the Arduino's electrically erasable programmable read-only memory (EEPROM). The Arduino, shields, and input BNC connectors are housed within a 3D printed enclosure. Below we describe each of these components in more detail.

#### 2.3.1. Servo shield

The servo shield is a 16-channel PWM shield with an Inter-Integrated Circuit (I2C) interface (Adafruit Product ID: 1411). For servo position updates, commands are sent from the Arduino to the shield over the I2C bus. In between updates, the shield holds the servo positions and does not require Arduino involvement. Several shields can be stacked if more than 16 channels are required (though some changes in case design and software would be required). If a display shield is used on top of the servo shield, the required headers limit the space for servo connectors (see build instructions). The Arduino is not able to supply enough power for the use of several servos simultaneously, so the servo shield utilizes a separate 5V power supply. To provide enough peak current for the simultaneous movement of several servos, place for a storage capacitor is provided on the shield circuit board – its capacity should be matched to the expected number of servos utilized.

*2.3.2. Display shield*

If the use of a display shield is desired, we provide two options: an LCD screen with buttons or a touch screen.

    The liquid-crystal display (LCD) shield (Adafruit Product ID: 772) contains a 16x2 character LCD and push buttons. On the display, the first line displays the shutter name, the second line the shutter status (open, closed, or inactive). The up/down buttons cycle through the selected shutter, the left/right buttons change the state (left for close, right for open, an optional timeout sets the shutter to an inactive state).

    The touchscreen shield (Adafruit Product ID: 1947) contains a 2.8", 240x320 pixel thin-film-transistor (TFT) LCD with capacitive touch sensing. The display portion uses the Arduino's Serial Peripheral Interface (SPI) bus, the touch sensing the I2C bus. Each line displays the shutter label in the center, and touch areas ("buttons") for opening and closing the shutter on either side. For ease of touching the correct buttons, the number of lines is currently restricted to four. However, if more shutters are desired, smaller fonts could be chosen (at the expense of smaller buttons) or line scrolling by swiping motions could be implemented.

*2.3.3. Digital input*

To control the shutters with an electrical signal, we implemented an interrupt-based monitoring routine of some of the Arduino digital input ports. The ports are configurable as 0 – 5V (TTL) inputs with (optional) input pull-up resistors that also allow the use of simple mechanical single-pole switches. The current case and software allow for four control lines, but this can be easily extended. Each shutter can be mapped to any (or no) control line and each control line can control several shutters. Note that while the inputs can be used as an interlock signal for laser safety applications, the RC servos keep their position in case of a power failure and hence do not provide a fail-safe closed state.

*2.3.4. Serial Communication*

Serial communication via the Arduino's USB port allows for programming of the shutter parameters, controlling the shutters, and checking the shutter status. Communication is done by sending three-character ASCII commands (with following command parameters, if applicable). A list of commands is provided in Table 1. The command format and example responses for the serial commands are listed in Table 2.

| Command | Description |
|---|---|
| *IDN? | Gets ID_STRING |
| GTI | Gets elapsed time in ms |
| GND | Gets the number of defined shutters |
| GSTx | Gets the state of shutter number x |
| GDLx | Gets the label of shutter number x |
| GTDx | Gets the stored transit delay of shutter number x in ms |
| GPRx | Gets all the stored parameters for shutter number x |
| SSTx | Sets the state of shutter number x |
| SSPx | Sets the current servo position of shutter number x |
| SPRx | Sets all the parameters stored for shutter number x |
| CLR | Clear the shutter parameters |
| SAV | Save the current shutter parameters to EEPROM |

**Table 1.** List of serial commands.

| Command | Example responses from the Arduino |
|---|---|
| *IDN? | "Arduino Uno Shutter 4.0" |
| GTI | "TI=434335" |
| GND | "ND=3" |
| GST0 | "ST=1" <br> Note: 0 -> closed, 1 -> open, 2 -> manual position set, -1 -> inactive |
| GDL1 | "DL=ShutterLabel2" |
| GTD0 | "TD=500" |
| GPR1 | "PR1,11,2,110,130,800, ShutterLabel2". <br> Format: <br> PRx,PWMChannel,digInputChannel,posOpen,posClosed,transitDelay_ms, Label |
| SST0,1 | Opens shutter number 0 <br> Returns "OK" if successful, an error otherwise. <br> Note: 0 -> close, 1 -> open |
| SSP0,200 | Sets the servo number 0 to position 200 <br> Returns "OK" if successful, an error otherwise. |
| SPR1,11,2,110,130,800, ShutterLabel2 | Sets the parameters for servo number 1. <br> Returns "OK" if successful, an error otherwise. <br> Format: see GPRx <br> Note: if x=-1 then a new shutter is added after the existing shutters |
| CLR | Returns "OK" |
| SAV | Saves only if at least one shutter is defined. <br> Returns "OK" if successful, an error otherwise. |

**Table 2.** Format and example responses for the serial commands.

Additional notes:
- For the communication to the Arduino, line termination is a line feed ('\n', LF, 0x0A) by default, but can be changed to a carriage return (CR, '\r', 0x0D). The response from the Arduino is the standard CR/LF ("\r\n").
- For commands that address a specific shutter (e.g. GSTx) the shutter number x has a zero-based index (0->first shutter, 1->second shutter). The same applies to the input control lines.
- Another SPR example: "SPR-1,3,-1,255,315,400,Name1" adds a new shutter labelled "Name1" after the existing shutters. The new shutter uses the servo shield channel 3, is not controlled by input lines, and has an open/close position of 255/315, and a transit delay of 400 ms.

*2.3.5. Parameter storage*

Each shutter is associated with a range of parameters:
- "PWMChannel" is the assigned hardware PWM channel number of the servo shield. Range: 0 to 15.
- "digInputChannel" is the input control line that controls the shutter state. Range: 0 to 3 and -1 (not controlled).
- "posOpen" and "posClosed" are the servo positions (PWM value) corresponding to the open/closed position.
- "transitDelay_ms" is the delay in ms that the shutter requires to fully open/close. This value is not used by the Arduino controller, but simply stored and can be returned upon request to implement wait times in a control sequence. This should be measured experimentally for each shutter.
- "Label" is the label displayed on the display. By default, this is limited to 7 characters (to fit on the touch screen display) but this can be extended in the configuration file.

These shutter parameters are stored in the EEPROM of the Arduino to retain their values after an Arduino reset.

*2.3.6. Arduino control modules and sequence*

The shutter control in the Arduino is split into modules, which can be utilized independently: the serial communication module, the display module, the control input module, and the idle check module. In the Arduino main loop, these modules (if utilized) are repeatedly queried for change requests.

The optional *serial communication module* handles serial communication between the Arduino and a computer through a USB connection.

For the optional *display module*, the LCD or TFT module can be utilized. Either will display the status of the shutter and let the user change it. A debounce mechanism is included for either device to avoid accidental multiple button presses. After a period of inactivity, either screen can dim and can be brightened again by any touch (for the TFT) or button press (for the LCD).

In the *control input module*, the control input lines are mapped to the Arduino's pin change interrupt mechanism. Even though interrupts can suspend all other Arduino activity when called, we decided to simply queue the state changes to be handled in the main loop. Given the relatively slow mechanical response time of an RC servo, the much more involved handling within the interrupt routine would not provide a noticeably improved response time. As in the display module, a debounce mechanism is included to avoid rapid erroneous state change requests (for example with a mechanical switch).

The *idle check module* (if enabled) checks when the controller last received a state change request and disengages the servo motors if an idle time has been exceeded. This can allow the user to move the shutter positions manually, which is only possible when the servos are disengaged. We found this capability to be convenient especially during optics alignment, where the shutter controller always seemed to be just out of easy reach.

*2.3.7. Library*

We provide a library for serial communication with the Arduino in both C and Python. The provided functions in the libraries handle the low-level serial communication and provide easy-to-use access functions. Both libraries utilize the Virtual Instrument Software Architecture (VISA) standard and provide wrapper functions (e.g. to set the shutter parameters or to open/close the shutters).

In addition to these libraries, we provide an example graphical user interface (GUI) in python, based on the tkinter library. Even though NI LabWindows/CVI is a commercial program (not freely available), it is a C IDE that is used in many labs (including ours) and as a convenience we provide the source to build a GUI using this platform.

## 3. Design files summary

| Design file directory name | File type | Open source license | Location of the file |
| --- | --- | --- | --- |
| Servo Mounts | CAD | *CERN-OHL-W-2.0* | Mendeley Data |
| Enclosures | CAD | *CERN-OHL-W-2.0* | Mendeley Data |
| Arduino Code | Source | *CERN-OHL-W-2.0* | Mendeley Data |
| C Library | Source | *CERN-OHL-W-2.0* | Mendeley Data |
| C GUI | Source | *CERN-OHL-W-2.0* | Mendeley Data |
| Python Library | Source | *CERN-OHL-W-2.0* | Mendeley Data |
| Python GUI | Source | *CERN-OHL-W-2.0* | Mendeley Data |

The *Servo Mounts* directory contains CAD files (both STEP and STL files) to 3D print mounts for attaching the RC servos to an optical post. Included files: ServoPostMount, ServoPostMount_Small

The *Enclosures* directory contains CAD files (both STEP and STL files) to 3D print an enclosure for the shutter controller. Included files: Enclosure_Bottom, Enclosure_Top_TFT, Enclosure_Top_LCD, Enclosure_Buttons_LCD

The *Arduino Code* directory contains the C source code for the Arduino shutter controller. Included files: ShutterDriverUniversal.ino, Common.h, SerialComm.cpp, SerialComm.h, LCD.cpp, LCD.h, TFT.cpp, TFT.h, Parameters.cpp, Parameters.h, DigInput.cpp, DigInput.h

The *C Library* directory contains source files for the C library that handles communication with the shutter controller. Included files: ArdShutter.c, ArdShutter.h

The *C GUI* directory contains source and resource files to build a test GUI to configure and test the shutter controller. Depends on the C library above and needs NI LabWindows/CVI to be installed. Included files: ArdShutterTest.c, ArdShutterTest.h, ArdShutterTest.uir

The *Python Library* directory contains source files for the python library that handles communication with the shutter controller. The logging level can be adjusted to include informational messages for debugging. Included files: ard_shutter.py

The *Python GUI* directory contains source files to build a test GUI to configure and test the shutter controller. Depends on the python library above. Included files: ard_shutter_test.py, ard_shutter_panel.py

## 4. Bill of materials summary

| Designator | Component | Number | Cost per unit - currency | Source of materials | Material type |
|---|---|---|---|---|---|
| *Controller* | *Arduino Uno* | 1 | $27.60 | Arduino.cc | Semiconductor |
| *Controller* | Servo shield | 1 | $17.50 | Adafruit | Semiconductor |
| *Controller* | LCD shield | 0 or 1 | $19.95 | Adafruit | Semiconductor |
| *Controller* | TFT shield | 0 or 1 | $44.95 | Adafruit | Semiconductor |
| *Controller* | Arduino power supply | 0 or 1 | $8.95 | Adafruit | Semiconductor |
| *Controller* | Servo shield power supply | 1 | $14.95 | Adafruit | Semiconductor |
| Controller | Various electronics | | ~$10 | Adafruit | Semiconductor |
| Shutter | Servo | >1 | ~$5-$35 | Varies | Semiconductor |
| Shutter | Various screws, metal pieces | | ~$10 | Varies | Metal |

For example, a shutter controller with four shutters using large RC servos, an LCD panel, and serial control would require an Arduino, a servo shield, an LCD shield, a servo power supply, four servos, and various electronics. The cost for this shutter system would amount to about $120.

## 5. Build instructions

General safety notice:
The assembly of the shutters and shutter controller involves 3D printing, mechanical assembly, electrical wiring, and soldering. To prevent damage to the electrical components, test power supply voltages before wiring cables to the Arduino and/or servo shield. All usual safety precautions should be taken when working with electronics or while soldering.

## 5.1 Shutter

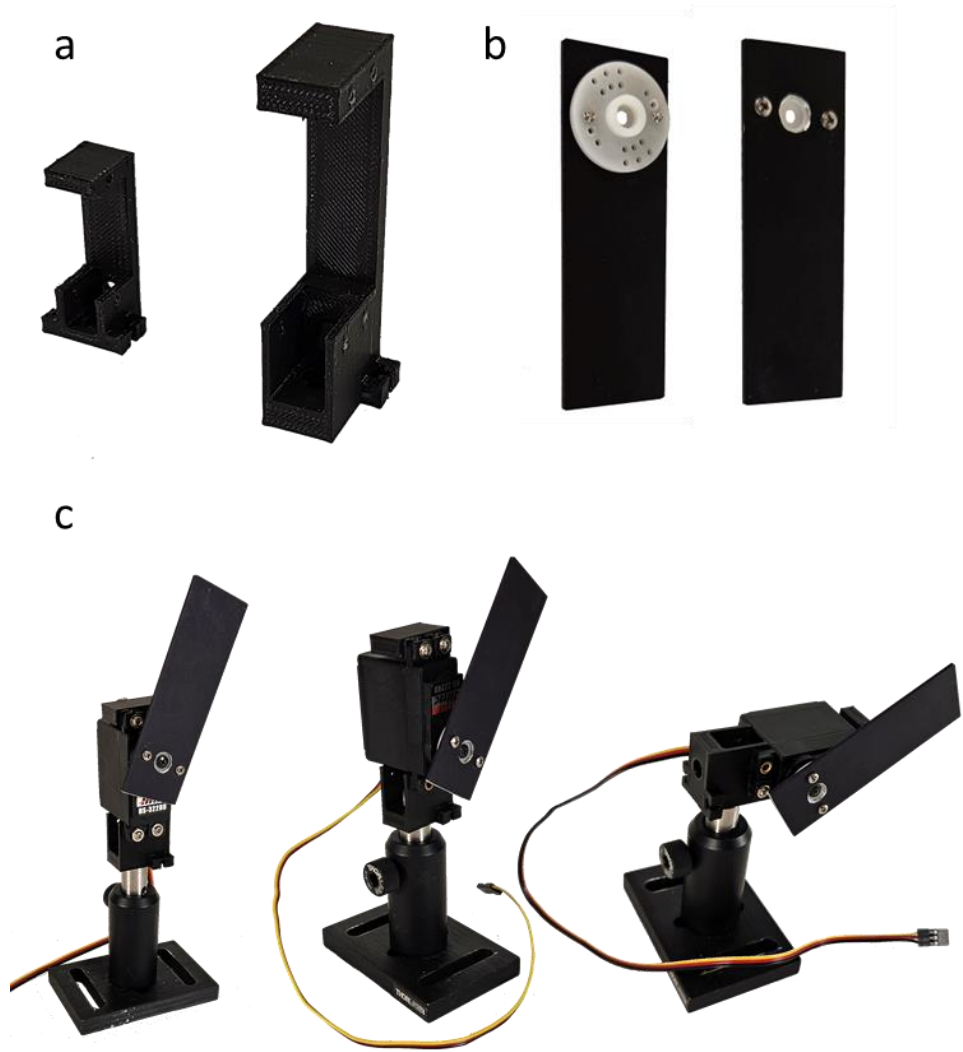

**Fig. 3.** Mounting of the servo and the shutter. Small and large servo mounts (a), back and front view of the mounted blade (b), and shutters mounted vertically (axis up and axis down) and horizontally (c).

Two sizes of shutter mounts (small and large) are provided for two common sizes of shutters (Fig. 3a). Small tabs are provided to fix the servo cables, if desired. For the blade, we drill a central hole large enough to clear the ledge on the mounting horn, and two small holes to attach it to the horn with self-tapping screws (Fig. 3b). For mounting, the servo slides in the U-shaped opening in the mount and is secured by four self-tapping screws (usually provided with the servo). Depending on the servo, rubber grommets are provided to minimize vibration transmitted from the servo to the mount. The servo can be mounted with its axle near the post mounting screw holes or opposite, depending on the light path requirements. (Fig. 3c). Finally, the assembled servo mount can be fixed with socket head screws (8-32 or M4 for the small mount, ¼-20 or M6 for the large mount) to a post.

## 5.2. Controller hardware

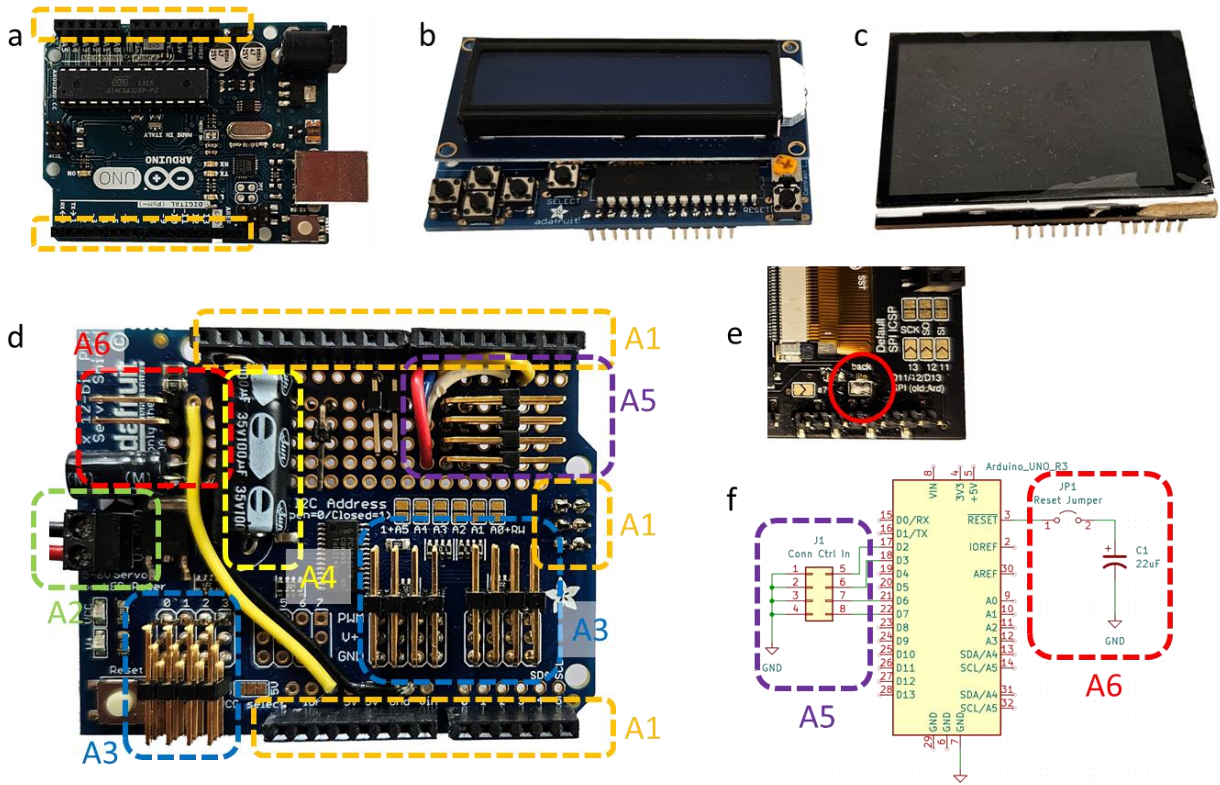

**Fig. 4.** Electronic modules: Arduino (a), LCD shield (b), TFT shield (c), and servo shield (front (d) and back(e)). Areas on the servo shields marked with color are: headers (A1), servo power connector (A2), right-angle RC servo connectors (A3), servo power supply capacitor (A4), right-angle control input connector (A5), and reset bypass capacitor and jumper (A6). Simplified electrical schematic (f), indicating electrical hookup in areas A5 and A6.

*5.2.1. Arduino*

In order to mount the servo shield, a set of female headers need to be soldered into the two rows of pins (Fig. 4a).

*5.2.2. Servo shield*

Technical details for this shield are provided by Adafruit [24]. To accommodate the display shield, two rows of female through headers and the upside-down SPI female through header need to be soldered into the servo shield (Fig. 4d, A1). A 2-pin screw terminal is soldered onto the shield for the power supply cable (A2), while the other end of the cable is attached to a power supply connector. The use of a cable connector reduces the risk of accidentally plugging the servo power supply into the Arduino power connector and vice versa (both power supplies have the same connector but a different voltage). Leaving all solder pads for the I2C address open results in the default address of 0x40. Because the display shield sits on top of the servo shield, use of angled servo connectors is necessary (A3), unless the cables are soldered directly into the board. We face the first connector outward for ease of access. The remainder of the connectors (if installed) need to face inwards because of the installed headers. Note that installing the header in the opposite direction reverses the order of the servo connector pins (GND on the top vs GND on the bottom). A capacitor to provide surge current for the servos can be installed on the board (A4); the value depends on the expected number of servos operated (see instructions on the Adafruit site). For the control input, we soldered another angled connector onto the board and connected one side to the microcontroller pins (PCINT18, PCINT19, PCINT22, and PCINT23 (Arduino pins D2, D3, D6, and D7; note that D4 and D5 are used by the TFT shield) and the other side to a common ground (see area A5). Finally, we connected a capacitor in series with a jumper to the reset pin of the Arduino (A6). This is a peculiarity when using the Arduino with the VISA library, where session initialization toggles the DTR line, which resets the Arduino. Connecting a capacitor between reset and ground suppresses this line toggle and allows for opening the serial port without reset [25]. During firmware programming of the Arduino, the capacitor needs to be disconnected by removing the jumper. Finally, on the bottom side of the shield there is a solder pad for the backlight of the screen (labelled back lite #5) – if screen dimming is to be enabled, this solder pad needs to be shorted with a dab of solder (see Fig. 4e).

*5.2.3. LCD shield*

The display shield is the topmost shield; hence, only short male headers are required (not stackable headers). Assembly instructions are provided by Adafruit [26]. No hardware I2C address selection is required.

*5.2.4. TFT shield*

The display shield is the topmost shield; hence, only short male headers are required (not stackable headers). Assembly instructions are provided by Adafruit [27]. No hardware I2C address selection is required.

*5.2.5. Cables*

The cables supplied with the servo are likely to be too short for typical use and extension cables are required. RC servo connector kits are available from online retailers that contain extension cables, connectors, and (if desired) a crimp tool for the connectors. To reduce noise pickup from the digital PWM signal by other electronics, a shielded cable (3 conductors + braided shield) can be used instead of the ribbon cable. In this case, connect the braided shielding to GND on the Arduino side of the cable. For the control inputs we use a short, stranded wire to the BNC connectors in the enclosure.

*5.2.6. Power supplies*

The Arduino can be powered through the USB port. If no USB connection is used (e.g. as a stand-alone shutter controller) a standard 9 V, 1 A wall-mount power supply is sufficient. For the servo shield, a 5 V power supply is recommended. The current rating depends on the number and type of servos that are being used (we use a 3 A supply for 4 servos).

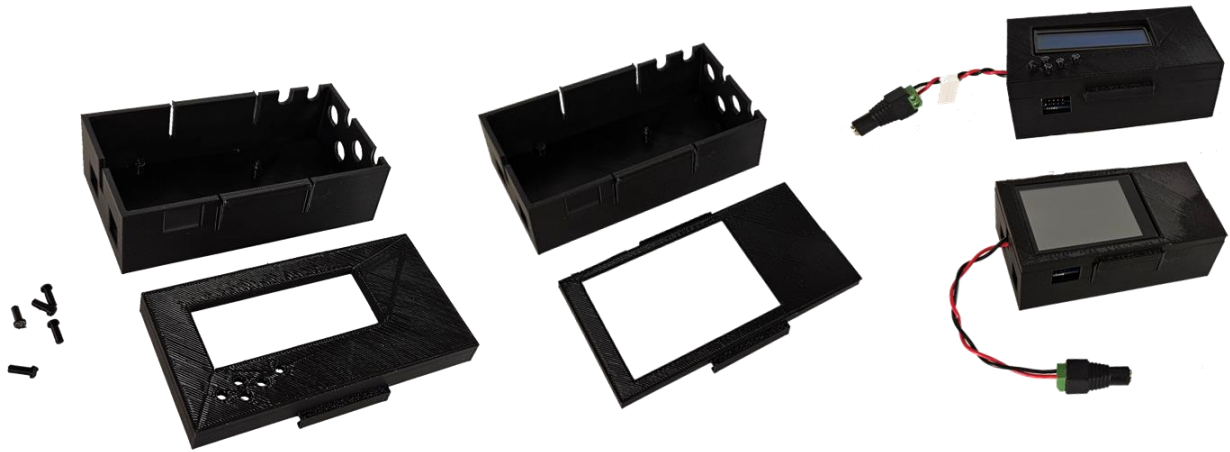

**Fig. 5.** Pictures of the enclosure for a controller in parts (left) and assembled (right).

### 5.2.7. Controller enclosure

The Arduino controller (incl. shields) is enclosed in a 3D-printed box. We provide a design for a controller with a servo shield and an LCD or a TFT shield. The enclosure is printed in two sections that are latched together. The bottom section has standoffs and guiding pins to position the Arduino and shields, D-shaped cutouts for the BNC connectors (used for the control lines), cutouts for power supply and USB cables, a rectangular cutout for up to 4 servo connectors, and cutouts for additional servo cables. The bottom section can be used for either the LCD or TFT display, whereas the top section is specific to the display type. Both top designs have cutouts for the display; the LCD model requires the insertion of small pins for the push buttons (the flared ends are inside the box to prevent them from falling out - assembly is easiest with the enclosure turned upside down). Tabs on the outside of the enclosure are provided for mounting on an optical table.

### 5.3. Controller software

### 5.3.1. Arduino code

Before the shutter can be used, the Arduino code needs to be compiled and uploaded. Opening the main "ShutterDriverUniversal.ino" in the Arduino IDE will open all associated source files as well. The code is split into several modules (a C++ file with a corresponding header file each). Each module has a compiler define "SERIAL_DEBUG" that can be set to 1 to receive status and warning messages via the serial monitor for debugging purposes. For normal operation these should be set to 0. The main module ("ShutterDriverUniversal.ino") contains the main Arduino setup and loop functions. Both functions instantiate and access objects/functions from the other modules. The communications module ("SerialComm.cpp") handles the communication over the serial (USB) port. The display modules ("LCD.cpp" and "TFT.cpp") handle the display and user input, the parameters module ("Parameters.cpp") the parameter storage in the EEPROM, and the digital control module ("DigInput.cpp") the shutter control via the control lines.

For convenience, the user-adjustable parameters are in the file "Common.h". The most important step is the inclusion of the parts of the program that are needed by uncommenting the compiler defines "SERIALCOMM", "DIGINPUT", "DISPLAY_TFT", and "DISPLAY_LCD" (at most one display can be defined). For example, if no control lines are used to control the shutters, the "DIGINPUT" define should be commented out so it will not get included during the compilation. The file "Common.h" also contains several useful constants that determine the program behavior, e.g. the maximum number of shutters allowed, the width of the borders on the TFT, the servo control frequency, the serial baud rate, and many others.

To compile the source code in the Arduino IDE, several libraries need to be included (the help menu in the IDE provides a link to library install instructions): the Adafruit PWM Servo Driver library for the Servo Shield, the Adafruit RGB LCD Shield library for the LCD Shield, and the Adafruit FT6206 library for the TFT Shield. Make sure to allow installation of dependent libraries by the Arduino IDE. After successful compilation, the Arduino code needs to be uploaded via the IDE (if the reset capacitor was installed, make sure to remove the jumper).

### *5.3.2. C library*

The C library handles low-level serial communication and provides easy-to-use wrapper functions (e.g. to set the shutter parameters or to open/close the shutters). The library utilizes the Virtual Instrument Software Architecture (VISA) standard, which needs to be installed on the system. Free VISA implementations (with installation instructions) are available from several companies, such as Tektronix [28], Keysight [29], or Rohde & Schwarz [30]. The supplied code was tested with NI-VISA [31] (which is, as of the time of writing, no longer free). The shutter C library (ArdShutter.c) only depends on the VISA library. The serial baud rate and termination character are defined in the same file. The header file (ArdShutter.h) only contains function prototypes for inclusion in other modules. The library currently only supports a single shutter controller and VISA handles are stored internally in the module. An instrument session needs to be established with ARD_ShutterInit before shutter commands can be issued. The session needs to be closed with ARD_Close when finished.

### *5.3.3. Python library*

The python library provides the same functionality as the C library above. This library, too, utilizes the Virtual Instrument Software Architecture (VISA) standard. For this library to work, the pyVISA library [32] needs to be installed. While pyVISA can use an installed VISA library from the beforementioned sources, it can also utilize pyVISA-py [33], an open-source, python-based VISA implementation. The supplied code was tested with pyVISA and pyVISA-py. The shutter controller is implemented as a class; its constructors and destructors handle instrument initialization and closing.

### *5.3.4. C GUI*

The C GUI tests the shutter functionality and uses the above C shutter library. To be compatible with the rest of the lab software, we used NI LabWindows/CVI (not free or open source - see the python GUI below for an open-source implementation). Though the source will not compile without the NI suite, the source can serve as example code on how to use the C shutter library functions.

### *5.3.5. Python GUI*

The python GUI tests the shutter functionality and uses the above python shutter library. The GUI is based on Tkinter, which comes built-in with most python installers.

## **6. Operation instructions**

General safety notice:
The shutters are designed to block light impinging on the shutter blade. Some heating of the blade and light scattering off the blade is expected and needs to be managed and monitored. When setting up and calibrating the shutters, adhere to all light (or laser) safety precautions.

**6.1 Initial setup**

Before the shutters attached to the controller can be used, the parameter settings need to be determined and uploaded to the controller. Upon boot, the Arduino reads saved parameter values from EEPROM, but does not move any servo motors unless directed by commands or user input. This gives the user the chance to safely program the parameters before first use (the EEPROM could initially contain random values). The following is the sequence for initial use (can be done using the library functions, with one of the GUIs, or directly with serial commands via the Arduino serial monitor):

1. Unplug the Arduino.
2. Connect the shutters to the servo shield and make a note of the port number used.
3. Plug the Arduino into a serial port.
4. Compile and download the Arduino code with the Arduino IDE.
5. Clear the parameter setting with the CLR command (see Note a in case of a stand-alone shutter controller).
6. For each shutter, send parameter values with the SPR command using "-1" as the shutter number – this value adds a new shutter to the parameter list. Make sure the port numbers match the ones from step 2. Use default values (e.g. 100,200,300) for the parameters openPos, closePos, and transitDelay (these values are calibrated in the next step.
7. Place the shutters in the respective beam path.
8. For each shutter, adjust the RC servo positions directly with the SSP command to find appropriate open and closed positions (note that if a position is unreachable, the servo horn can be attached at a different angle). Update the openPos and closePos parameters for each servo with the SPR command, this time using the respective shutter number (instead of the -1 used previously, remember that the index is zero-based). An easy way to do this is to get the parameters with the GPR command, change the positional values, and send the updated parameters back with the SPR command.
9. If the transit delay parameters are used (these do not affect the shutter operation and are just stored for queries) they need to be calibrated, for example with a photodiode in the light path. Once determined, update this value (like the update in step 6).
10. Check the parameter values (and order) individually.
11. Save the parameters to EEPROM with the SAV command.

Notes:

a) If no serial connection is used in the shutter controller, the parameters can be set when downloading the program to the Arduino (see Note b below).
b) In the unlikely event that random initial values in the EEPROM make the Arduino behave erroneously when first programmed and powered up (or when no serial connection is used for the shutter operation), we provide a routine "createDummyParameters" in the file "ShutterDriverUniversal.ino" that could be temporarily substituted for "_params.readFromEEPROM" in the setup portion of the Arduino to pre-set the parameters in the EEPROM (see instructions in the source code).

**6.2 Standard operation**

The shutter parameters are stored in the Arduino and are loaded at boot time. No calibration is needed during normal operation. The following features are enabled by default (but can be disabled in the source code):

- The screens dim after a period of inactivity (default 60 s, set LCD_DIM_PERIOD_S or TFT_DIM_PERIOD_S to zero to disable). Wake-up is through any button press or touch.
- For the LCD display, the up/down buttons cycle through the shutters, the left/right closes/opens them. For the TFT display, actions are button touches.
- The RC servos disengage (are movable by hand) after a period of inactivity (default 5 s, set IDLEINTERVAL_S to zero to disable this mode).

## 7. Validation and characterization

The performance of the shutter (opening/closing time, opening/closing delays) varies widely with use parameters, like the RC servo type, size of the blade, size of the light path, and relative positioning. We tested operation with two different RC servos: a small servo (Savox SH-0262MG) and a large servo (HiTEC HS-322HD). A laser diode (beam diameter approx. 2 mm in the direction of the shutter blade movement) was directed onto a photodiode and the output was monitored with an oscilloscope. The shutter was positioned such that the blade pivot point was approximately 3 cm away from the beam. The distance from the blade to the beam (in the open position) and from blade to the point where the beam hits the blade (in the closed position) was about 3 mm.

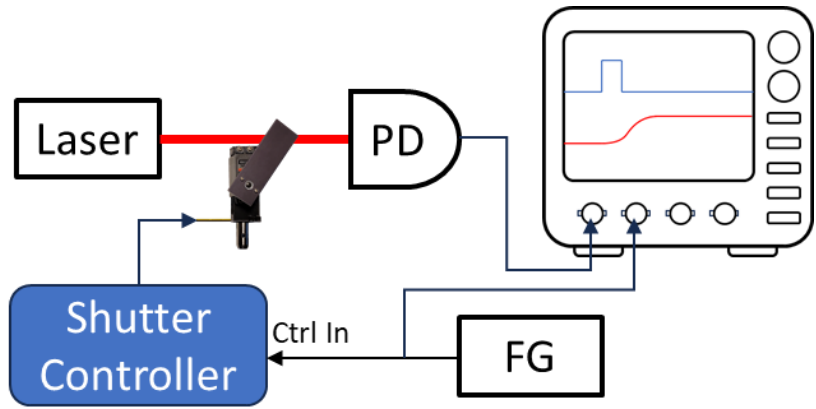

**Fig. 6.** Test setup to measure opening/closing time/delay. PD: photodiode, FG: function generator.

A TTL pulse from a function generator toggled the shutter (open/close) through the control input of the shutter controller and provided a reference for the oscilloscope. With this setup, we were able to measure the opening/closing and delay times. We define the opening time as the rise in photodiode signal from 20% to 80% full scale and the closing time as the falling time from 80% to 20%. The opening/closing delay is the time from the change in the control signal to the midpoint of the opening/closing signal.

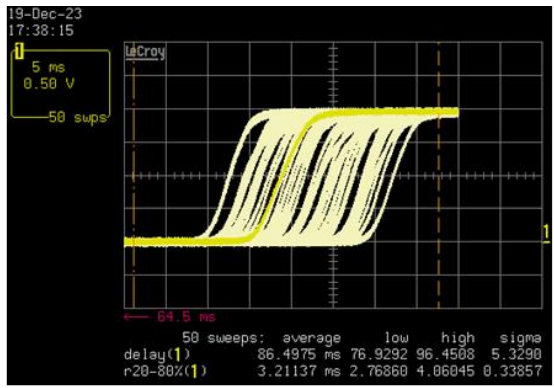 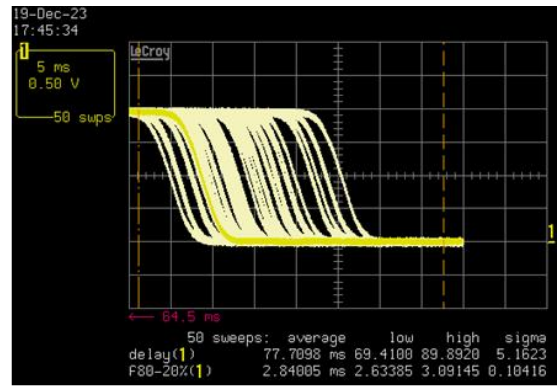
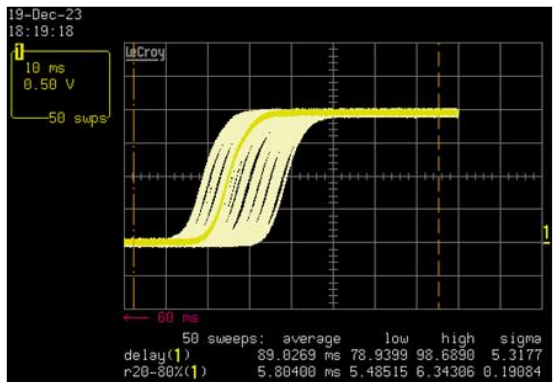 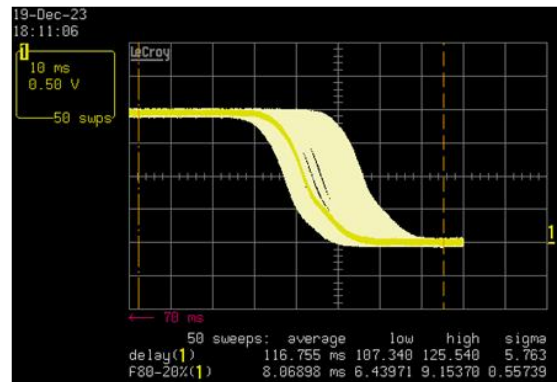

**Fig. 7.** Measured opening and closing performance in the test setup.

With the small servo (rated speed of 60 deg / 80 ms), we measured an opening time of (3.2 ± 0.3) ms and a closing time of (2.8 ± 0.1) ms. The opening/closing times agree qualitatively with the expected value of 5 ms according to Eqn. 1. Note that this approximate equation assumes a uniform (top-hat) beam and measures 0 to 100% transit times – our test beam is not a top-hat beam but falls of gradually from the center. The delay times depend strongly on the positioning of the shutter with respect to the beam. In our case, we measured delays on the order of 100 ms, 50 ms of which are caused by the software debounce routine on the control input. By disabling debouncing in the Arduino code (clean TTL control signals do not need debouncing) and minimizing the beam width and distance from the blade edge, delays below 50 ms are easily achievable. It is worth noting that the variation in the delay times (a range of about 20 ms) is consistent with the 50 Hz operating frequency of the PWM controller, dictated by the servos. Any update request from the PWM controller that occurs between the pulse repetition period (20 ms) is deferred until the next period.We also measured the timings for the larger shutter (rated speed of 60 deg / 190 ms). The opening/closing times are correspondingly longer, while the delay variation is similar (see Fig. 7).

## CRediT author statement

**Mathias Fischer:** Software, Validation, Investigation, Writing - Reviewing and Editing,
**Martin Fischer:** Conceptualization, Methodology, Software, Writing - Original draft preparation, Funding Acquisition, Supervision.

## Acknowledgments

This material is based upon work supported by the National Science Foundation under grant no. CHE-2108623, and by the Chan Zuckerberg Initiative DAF, an advised fund of Silicon Valley Community Foundation under grant no. 2021-242921. We also acknowledge Dr. David Grass for helpful suggestions for manuscript preparation.

# Appendix A: BOM details

## Controller

| Component | Number | Cost /unit | Source | Part number | Link | Comments |
|---|---|---|---|---|---|---|
| Arduino Uno R3 | 1 | $27.60 | Arduino.cc | Arduino A000066 | https://store-usa.arduino.cc/products/arduino-uno-rev3 | Arduino clones can be obtained (cheaper) from a variety of online sources |
| Servo Shield | 1 | $17.50 | Adafruit | Adafruit 1411 | https://www.adafruit.com/product/1411 | |
| Stacking headers | 1 | $1.95 | Adafruit | Adafruit 85 | https://www.adafruit.com/product/85 | Only required if a display shield is used |
| Pack of right-angle headers | 1 | $2.95 | Adafruit | Adafruit 816 | https://www.adafruit.com/product/816 | Only required if a display shield is used |
| LCD shield | 1 | $19.95 | Adafruit | Adafruit 772 | https://www.adafruit.com/product/772 | Optional |
| TFT shield | 1 | $44.95 | Adafruit | Adafruit 1947 | https://www.adafruit.com/product/1947 | Optional |
| 9V power supply for Arduino | 1 | $8.95 | Adafruit | Adafruit 63 | https://www.adafruit.com/product/63 | Only required if not connected to USB |
| 5V power supply for servo shield | 1 | $14.95 | Adafruit | Adafruit 1466 | https://www.adafruit.com/product/1466 | Amperage depends on the number of servos |
| Power adapter jack to screw terminal | 1 | $2 | Adafruit | Adafruit 368 | https://www.adafruit.com/product/368 | Used to connect the servo shield power supply to the servo shield board |
| BNC connectors | 0-4 | $3.76 | Digikey and others | Amphenol 31-10-RFX | https://www.digikey.com/en/products/detail/amphenol-rf/31-10-RFX/100642 | Only required if control lines are used |

The following items might be required, but these are common items in any electronics bench:
- Electrolytic capacitor (value depends on number of servos) - only required if several servos are used (see manuscript).
- Electrolytic capacitor (~22 µF) and a jumper - only required if serial communication resets Arduino upon port open (see manuscript).
- Some wire and solder accessories.

## Servos

A variety of servos are available that serve the purpose of a shutter. The ones we tested were:
- Small servo: Savox SH-0262MG. As of the time of manuscript submission, this item is discontinued. A replacement item is SH-0264MG at a price of $35.99 each. https://www.savoxusa.com/products/savsh0264mg-super-torque-metal-gear-micro
- Large servo: MG 996R at a price of about $5 each. 4-pack: https://www.amazon.com/4-Pack-MG996R-Torque-Digital-Helicopter/dp/B07MFK266B/

- Large servo: HiTEC HS-322HD at a price of about $15 each. https://www.amazon.com/Hitec-33322S-HS-322HD-Standard-Karbonite/dp/B0006O3XEA/
- If the cables of the servos are too short (very likely), extension cables can be made with a servo cable kit. A crimp tool makes assembly easier - this tool is included in many kits, for example: https://www.amazon.com/Female-Connector-Crimping-Compatible-Spektrum/dp/B0BGX7157Y/. Price: about $24 each
- For the blade, a scrap piece of anodized aluminum will work. We used a small section off a 1/8" thick, 1 ft strip: https://www.mcmaster.com/7083T11/ Price: $12.50/ft
- To attach the blade to the servo horn, most short #0 self-tapping screw will do, e.g. https://www.mcmaster.com/92470A018/ (Price: $7/pack of 50 screws)